\documentclass[a4paper]{article}

\usepackage{INTERSPEECH2021}
\usepackage[abs]{overpic}
\usepackage{xcolor,varwidth,relsize}
\usepackage{caption}
\usepackage{setspace}
\usepackage{soul}
\usepackage{nicefrac}
\newcommand{\numgenerations}{T}
\newcommand{\gennum}{t}
\newcommand{\dataset}{D}
\newcommand{\datasetdist}{\mathcal{D}}
\newcommand{\mdl}{\mathcal{F}}
\newcommand{\diffobject}{\mathcal{G}}
\newcommand{\loss}{\mathcal{L}}
\DeclareMathOperator*{\argmin}{argmin}

\makeatletter
\def\blfootnote{\gdef\@thefnmark{}\@footnotetext}
\makeatother

\title{Learning a Neural Diff for Speech Models}
\name{Jonathan Macoskey, Grant P. Strimel, Ariya Rastrow}
\address{Alexa Machine Learning, Amazon.com, USA}
\email{$\{$macoskey,gsstrime,arastrow$\}$@amazon.com }
\begin{document}

\maketitle
\begin{abstract}
As more speech processing applications execute locally on edge devices, a set of resource constraints must be considered.
In this work we address one of these constraints, namely over-the-network data budgets for transferring models from server to device.
We present neural update approaches for release of subsequent speech model generations abiding by a data budget.
We detail two architecture-agnostic methods which learn compact representations for transmission to devices.
We experimentally validate our techniques with results on two tasks (automatic speech recognition and spoken language understanding) on open source data sets by demonstrating when applied in succession, our budgeted updates outperform comparable model compression baselines by significant margins.
\end{abstract}

\section{Introduction}
\blfootnote{The first two authors contributed equally.}
With the expanding capabilities of smart devices and speakers, a recent trend has emerged to enable local speech processing for virtual assistants such as Amazon Alexa and Google Assistant \cite{JinyuLiRuiZhaoHuHu2019, McGraw2016}.
While local execution of speech models yields benefits such as reduced latency and enhanced reliability \cite{Sainath2020}, the edge-first architecture poses a series of challenges traditionally not faced for a server-centric one.
Prior work has emphasized the two most manifest constraints: limited compute resources which impacts latency \cite{He2018, Macoskey2021} and memory constraints for storage of large models locally \cite{Pang2018}.
However, another challenge for on-device speech which has yet to be studied to the same degree is imposing over-the-network, or \textit{over-the-air} (OTA), data budgets.
One of the primary benefits of a server-based speech architecture is the ease of deployment for new model generations or improvements which can occur near instantaneously and at low cost.
In industry settings, having regular updates remains a critical component of the machine learning flywheel:
enhanced model releases deliver more accuracy and functionality, which drives user adoption of the technology, which in turn leads to more data for building superior models.
Meanwhile, users are able to experience the added benefits or functionality immediately when new model generations are deployed to the server.
When speech models are executed locally, however, an individual who has multiple speech enabled devices (e.g. phone, smart speaker, vehicle) would require a download for each when a model is available.
Speech models can be large while an individual's network data plan might be restrictive (e.g. 1 GB of data monthly before higher data rates apply).
Particularly relevant for remote or developing regions, where network availability can be limited, smaller artifacts can also provide faster download times.

To complicate matters, another trend which has complimented that of on-device technology is the adoption of end-to-end neural architectures for speech \cite{Graves2012, Graves2014, Chan2016, Zhang2020}.
While often superior in performance, fully neural speech architectures can have a large footprint (120+ million parameters \cite{Narayanan2019}) which are expensive with respect to data transfer.
Furthermore, by their end-to-end nature, these architectures blur the roles of model components when compared to their traditionally decomposed counterparts.
For example, a conventional DNN-HMM ASR system with a disjoint acoustic model, pronunciation model, and language model might only require an update to a segment of just a single subsystem (e.g. adjusting a weighted finite state transducer).
Or, consider a two-phase spoken language understanding (SLU) system: ASR followed by natural language understanding (NLU) where only the NLU component might require an update for a new functionality release.
In such cases only the new NLU module, which can be highly compressed  \cite{Strimel2018, Mysore2020}, would need to be transferred.
Thus, the shift to end-to-end neural models motivates the need to develop holistic approaches for efficient edge model update methods addressing data budget constraints.
In this paper we consider the problem of incrementally updating models for evolving and expanding training data sets with minimal OTA transfer costs.
We refer to this here as \emph{neural diff learning}, analogous to a source code diff for a software patch.

Communication efficient model updating has been studied in several related contexts, one of the most prominent being with respect to data-parallel distributed model training.
There is a rich literature on effective gradient compression schemes \cite{Seide2014,Strom2015,Alistarh2017,Yujun2018,Spring2019} which are designed to limit the I/O bottleneck of communication and aggregation of gradient information applied to a model in training at each step of gradient descent.
In contrast, with our setting we aim to \textit{train} just a single, communication efficient artifact to apply to the prior model.
Our setting is also closely related to the modern continual learning (CL) and federated learning (FL) class of problems which both have broad bodies of work \cite{Lange2019,Sadhu2020, Mcmahan2017, Dimitriadis2020}.
CL typically addresses the problem of never-ending model adaptation over an evolving data stream with a particular focus on preventing \emph{catastrophic forgetting} of learnings acquired over no longer accessible data \cite{Li2018,Farajtabar2020}.
FL draws on ideas from both the CL and distributed training spaces to enable data-private, decentralized learning. Among a plethora of other challenges, FL often requires techniques for efficient two-way communication and synchronization for all client-devices in order to be practical at scale \cite{Kairouz}.
In our study here, we will retain a growing data set server-side and focus on server-to-device communication, though our techniques draw inspiration from these fields, particularly the regularization based methods of CL \cite{Li2018}.

Our contributions include formalizing our neural diffing problem; presenting two architecture agnostic approaches for minimizing communication cost for model updates; and applying and evaluating these techniques in speech settings for two tasks, ASR and SLU classification, on publicly available data sets.

\section{Problem Setting}
To encapsulate the notion of growing and evolving training data for a production voice-assistant system, we define a series of $\numgenerations$ data sets $\dataset_1 \subset \dataset_2  \subset \cdots \subset \dataset_\numgenerations$.
Each data set corresponds to a model $\mdl$ in sequence of model generations $\mdl_1, \mdl_2, \dots, \mdl_\numgenerations$ which are built successively on the data sets and deployed at regular intervals.

\subsection{Objectives}
In a traditionally unconstrained setting, we simply aim to train the best possible model over the given data with respect to a loss function $\loss$. 
In this scenario each model generation is defined approximately as 
\[
\mdl_\gennum^* \approx \argmin_{\mdl} \sum_{(x,y) \in \dataset_\gennum} \loss \left(\mdl_\gennum(x), y \right).
\]
For convenience we will hereafter denote $\loss_{\dataset}\left( \mdl \right)$ as an abbreviation for the loss of a model $\mdl$ over some dataset $\dataset$.

In the constrained setting we consider in this work, our goal now, broadly construed, is to train and deploy a series of models $\mdl_1, \dots, \mdl_\numgenerations$ where at each generation $\gennum$ we aim to minimize the predictive performance regret under the OTA budget when compared to the unconstrained ideal $\mdl_\gennum^*$. Namely,
 
\[
\text{regret}_\gennum = \loss_{\dataset_\gennum} \left( \mdl_\gennum \right) - \loss_{\dataset_t} \left( \mdl_\gennum^* \right)
\]
subject to the following constraint
\[
\Delta(\mdl_\gennum, \mdl_{\gennum+1}) \leq B,
\]
where $\Delta$ measures the diff size between subsequent model generations and $B$ is the OTA update budget per generation.
Last, we assume the device comes prepackaged with an initial generation model $\mdl_0$ out of the box or at initial application download.

\subsection{Data Set Evolution}
Our data sets consist of $(x,y)$ pairs where $x$ is a feature vector (e.g. audio features) and $y$ is a label for the instance (e.g. $y \in \left[ C \right]$ for a classification task such as speech commands, $y \in \Sigma^\ell$ for sequence labeling such as open vocabulary speech recognition).
We consider the evolution of the data set sequence under two settings.

\subsubsection{Setting 1}
With this evolution there is a global distribution $\datasetdist$ from where data at all times is sampled. However, as the training data becomes more abundant, i.e. $|\dataset_\gennum| \leq |\dataset_{\gennum+1}|$, the task and associated loss becomes more precisely defined.
We expect the model to improve performance on its task at each generation because it simply is exposed to more training data.

\subsubsection{Setting 2}
\label{sec:setting_2}
In this scenario, the evolution of the dataset shifts the task. Thus each generation has $\dataset_\gennum \sim \datasetdist_\gennum$ but $\datasetdist_\gennum \neq \datasetdist_{\gennum+1}$.
These different distributions could simply shift the sampling frequency of certain examples (e.g. a trending word or phrase) or be the introduction of a new $y$ class label (e.g. a new command now enabled with the release of a new functionality).
While still $|\dataset_\gennum| \leq |\dataset_{\gennum+1}|$, there is not necessarily an expectation that $\loss_{\dataset_\gennum}\left( \mdl_\gennum^* \right) \geq \loss_{\dataset_{\gennum+1}}\left( \mdl_{\gennum+1}^* \right)$ because the task difficulty could be different during the evolution.

\section{Diffing Approach}
Our general approach to  updating models using a low OTA budget will be to directly learn a small diff (or patch) to apply to the previous model generation over the expanded training data.
We do this by anchoring the previous model generation, then learning the diff object $\diffobject_{\gennum+1}$ as free weights
\[
\mdl_{\gennum+1} \approx \mdl_{\gennum} \oplus \argmin_{|\diffobject_{\gennum+1}| \leq B} \loss_{\dataset_{\gennum+1}} \left( \mdl_{\gennum} \oplus \diffobject_{\gennum+1} \right)
\]
where $\oplus$ is an operator expressing the application of the learned diff and $|\diffobject|$ designates the size of a diff $\diffobject$.
As a result, only $\diffobject_{\gennum+1}$ needs to be transferred to the local device where it can then be applied to $\mdl_{\gennum}$.

In this work, our implementation will view $\diffobject$ as a series of matrix masks each corresponding to a weight matrix of $\mdl$.
Each mask is applied through a simple additive or multiplicative operation (see below) to its matching anchored weight matrix during the forward pass of training.

The masks will be subject to a size budget (not dimension budget) bounding their footprint.
Below we detail two generic ways to learn budgeted diffs, one leveraging sparsity and another using weight hashing.
In a traditional model compression setting, these techniques have limits before noticeable accuracy degradation appears; however, when applied to diff objects instead of original model weights, our motivating intuition is that the diff objects can be compressed to a much higher degree by exploiting and reusing the information content of the prior model generation.
As a result, the methods can potentially surpass the observed raw model compression barriers.

\subsection{Sparse Diffing}
Many works have demonstrated that deep neural networks can be trained effectively using sparse representations \cite{Liu2015, Pang2018, Louizos2018, Zhu2018, Zhen2021}. 
Sparsity is attractive not only for reducing a model's inference operations but also its memory footprint.
In this work we apply sparsity to the additive weight masks of  $\diffobject$. 
Specifically, we use the pruning approach of \cite{Zhu2018} which presents a method to train to any desired sparsity by gradually ``zero-ing out weights'' of smallest magnitude during the training process. 
For a target sparsity $s_f$, initial pruning step $r_0$, and final pruning step $r_f$, we prune the $s_r$ fraction of diff weights of least magnitude according to schedule
\[
s_r = s_f \cdot \left( 1 - \left( 1 - \frac{r - r_0}{r_f - r_0} \right)^3 \right).
\]

\begin{figure}
\centering
\begin{overpic}[width=0.9\linewidth]{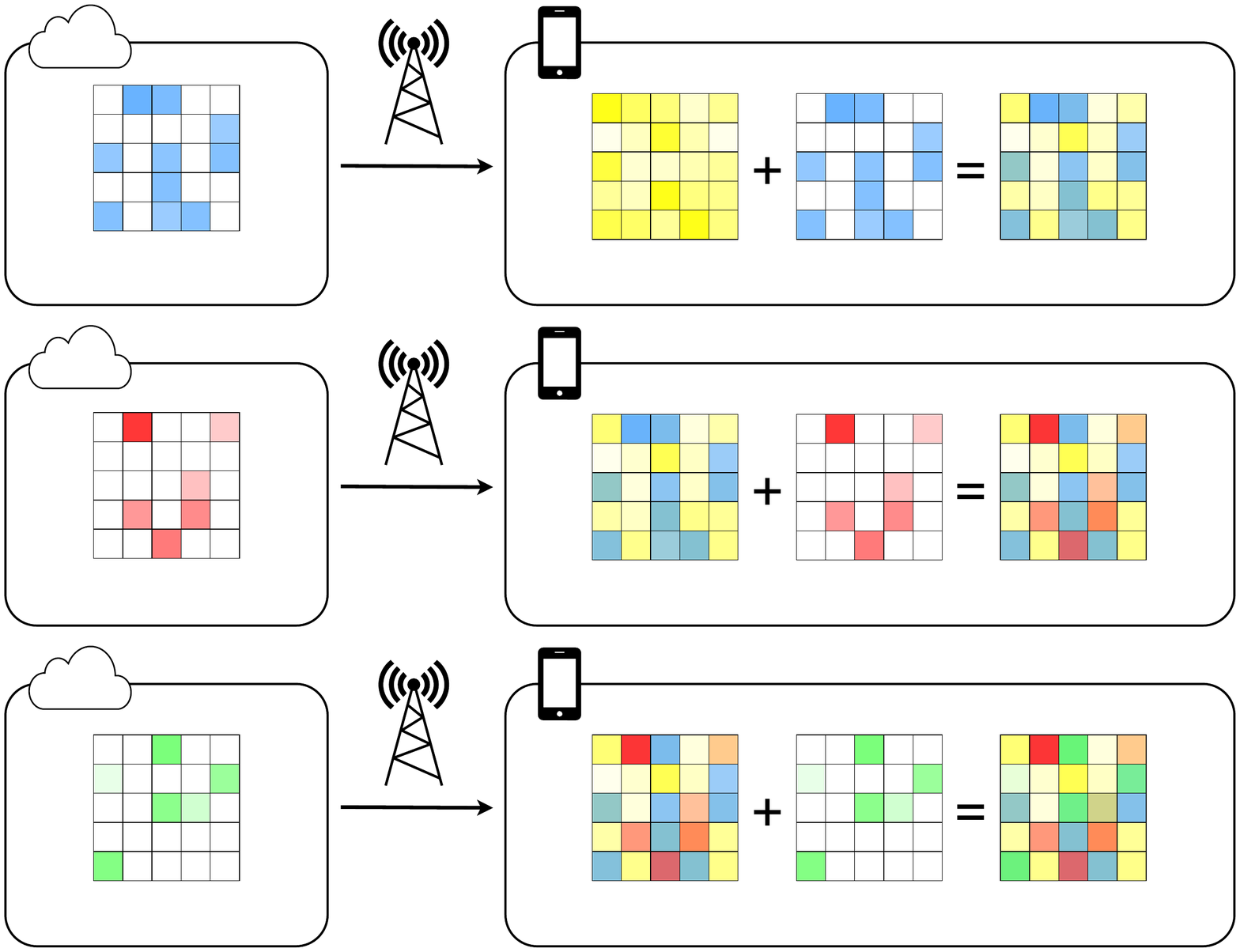}
\put(33.8,106.5){\parbox{0.0\linewidth}{\scalebox{0.85}{$\diffobject_1$}}}
\put(105,106.25){\parbox{0.0\linewidth}{\scalebox{0.85}{$\mdl_0$}}}
\put(135,106.5){\parbox{0.0\linewidth}{\scalebox{0.85}{$\diffobject_1$}}}
\put(163.75,106.25){\parbox{0.0\linewidth}{\scalebox{0.85}{$\mdl_1$}}}

\put(33.8, 60){\parbox{0.0\linewidth}{\scalebox{0.85}{$\diffobject_2$}}}
\put(105,59.75){\parbox{0.0\linewidth}{\scalebox{0.85}{$\mdl_1$}}}
\put(135,60){\parbox{0.0\linewidth}{\scalebox{0.85}{$\diffobject_2$}}}
\put(163.75,59.75){\parbox{0.0\linewidth}{\scalebox{0.85}{$\mdl_2$}}}

\put(33.8,13.75){\parbox{0.0\linewidth}{\scalebox{0.85}{$\diffobject_3$}}}
\put(105,13.5){\parbox{0.0\linewidth}{\scalebox{0.85}{$\mdl_2$}}}
\put(135,13.75){\parbox{0.0\linewidth}{\scalebox{0.85}{$\diffobject_3$}}}
\put(163.75,13.5){\parbox{0.0\linewidth}{\scalebox{0.85}{$\mdl_3$}}}
\end{overpic}
\caption{Sparse diffing procedure. Each model $\mdl_t$ has learned sparse weight masks $\diffobject_{\gennum+1}$ applied to arrive at the next generation model $\mdl_{\gennum+1}$. \vspace{-4.5mm}}
\label{fig:diff_sparsity}
\end{figure}

Since we prune diff weights and not model weights, the updated model on the device remains fully dense after the application of the diff.
A major strength of this approach is that as mask weights are pruned, the resulting model weights are not ``zeroed-out'' as they would be when sparse pruning a model directly.
Rather, the weights are simply substituted with the prior generation's weights.
So while prior works show that a model performance begins to degrade at a particular level of sparsity (\cite{Pang2018, Zhen2021} show $\sim50$\% for their speech models), we circumvent this compression barrier by retaining a fully dense model for inference.
Furthermore, the sparse diffing process has the ability to dynamically adjust sparsity structure generation over generation (see Figure \ref{fig:diff_sparsity}).
While a subset of non-zero diff weights are used for a particular update, an entirely different set may be selected during the next generation.
Last, observe that the approach remains generic. 
It makes no assumptions about the roles of model components and does not restrict which weights of the neural network can be updated.

\subsection{Hash Diffing}
\label{sec:hash_diff}
Weight hashing is another technique which has been shown effective in compressing large neural networks \cite{Chen2015}.
Here we adapt the technique for the diffing setting by using it to compress the weight masks of a diff object.
Given a weight matrix $W_\gennum$ of $\mdl_\gennum$  we train an array of diff weights $A_{\gennum+1}$ so that
\[
W_{\gennum+1}^{(i,j)} = W_{\gennum}^{(i,j)} + |W_{\gennum}^{(i,j)}| \sum_{k=1}^K A_{\gennum+1}\left[ h_k(i,j) \right],
\]
where $W_{\gennum+1}^{(i,j)}$ is the $(i,j)$th entry of the updated weights for model generation $\mdl_{\gennum + 1}$ and $h_k$ is one of $K$ hash functions randomly hashing weight positions into locations of $A$. 
The size of $A$ is chosen to match one's budget $B$, which is much smaller than the size of $W$.
As a result, there is a high rate of collisions leading to heavy randomized weight sharing.
Because of the high collision rate, we add the scaling factor $|W_{\gennum}^{(i,j)}|$ \footnote{This is a unique aspect in the diffing case over the original method because we leverage predefined weights from prior generations}.
This factor essentially changes what would be an additive update operation into a multiplicative one, with the added benefit that when expressed this way allows the trainable weight distribution to be centered at zero (i.e. identity is 0 instead of 1).
While not strictly necessary, we found empirically that magnitude scaling was more effective than plain additive weight updates because it is much easier to shift a pair of colliding weights proportionally the same amount rather than absolutely (e.g. consider a large weight colliding with a smaller one).

While weight hashing is a compelling method capable of delivering significant compression, it does not appear to be the common method of choice for representing neural models for real-time inference applications on-device.
Among other reasons, the limited adoption can be attributed to the random access patterns in memory the hash functions produce when accessing weights, while modern system hardware, on the other hand, is designed to support efficient spatial locality for bulk matrix operations at high speed.
Interestingly, in this setting where the hashing is used only to represent the diff object during transmission before it is decompressed and applied to the previous model on device, we do not encounter this issue.
In addition, as \cite{Shi2009,Chen2015} suggest, the use of multiple hash functions ($K > 1$) can provide extra modeling capacity to the compression but come at the cost of multiple memory accesses and addition operations to reconstruct a single weight.
Again, in the diffing instance, this reconstruction work can occur offline before being applied to the dense model used at inference time.
Last, generation over generation, we have the ability to reseed our hash functions to prevent the same weight positions from always colliding.
This adaptation is analogous to the sparse method selecting different non-zero diff weights each generation.

\setul{0.7ex}{}
\section{Experimental Results}
\begin{table*}[]
\caption{Comparison of an ideal model's performance baseline (BL) against baseline model compression (BLC) and incremental diff application (Diff) under different OTA budgets compression ratios (CR) at each model generation.}
\vspace{-2.5mm}
\renewcommand{\arraystretch}{1.2}
\resizebox{\textwidth}{!}{%
\begin{tabular}{@{}c|clclcc|ccccccccccc@{}}
\toprule
               & \multicolumn{6}{c|}{\textbf{LibriSpeech (WER)}}                                                                                      & \multicolumn{11}{c}{\textbf{Fluent Speech (Classification Acc.)}}                                                                                                                                                    \\ \midrule
               & \textbf{}  & \multicolumn{1}{c}{\textbf{}}   & \multicolumn{2}{c}{\ul{\mbox{\hspace{2.5mm}}Sparse\mbox{\hspace{2.5mm}}}}                & \multicolumn{2}{c|}{\ul{\mbox{\hspace{2.5mm}}Hash\mbox{\hspace{2.5mm}}}} & \textbf{}  & \textbf{}   & \multicolumn{4}{c}{\ul{\mbox{\hspace{8.5mm}}Sparse\mbox{\hspace{8.5mm}}}}                                                                & \multicolumn{4}{c}{\ul{\mbox{\hspace{10.5mm}}Hash\mbox{\hspace{10.5mm}}}}                       & \ul{\mbox{\hspace{0.5mm}}Static\mbox{\hspace{0.5mm}}} \\
\textbf{CR} & \textbf{}  & \multicolumn{1}{c}{\ul{\mbox{\hspace{0.5mm}}1\mbox{\hspace{0.5mm}}}}           & \multicolumn{2}{c}{\ul{\mbox{\hspace{6.5mm}}10\mbox{\hspace{6.5mm}}}}                        & \multicolumn{2}{c|}{\ul{\mbox{\hspace{6.5mm}}10\mbox{\hspace{6.5mm}}}}          &            & \ul{\mbox{\hspace{0.5mm}}1\mbox{\hspace{0.5mm}}}            & \multicolumn{2}{c}{\ul{\mbox{\hspace{6.5mm}}20\mbox{\hspace{6.5mm}}}}                              & \multicolumn{2}{c}{\ul{\mbox{\hspace{6.5mm}}40\mbox{\hspace{6.5mm}}}}                  & \multicolumn{2}{c}{\ul{\mbox{\hspace{6.5mm}}20\mbox{\hspace{6.5mm}}}}    & \multicolumn{2}{c}{\ul{\mbox{\hspace{6.5mm}}40\mbox{\hspace{6.5mm}}}}    & \ul{\mbox{\hspace{0.5mm}}inf\mbox{\hspace{0.5mm}}}         \\
               & hours & \multicolumn{1}{c}{BL} & BLC & \multicolumn{1}{c}{Diff} & BLC     & Diff    & \# dom. & BL & BLC             & Diff               & BLC & Diff               & BLC & Diff & BLC & Diff & $\text{BL}_\infty$ \\ \midrule
\textbf{Gen 0} & 480        & 9.93                            & -            & \multicolumn{1}{c}{-}          & -                & -             & 3          & 97.5        & -                        & -                        & -            & -                        & -            & -          & -            & -          & -           \\
\textbf{Gen 1} & 576        & 9.37                            & 9.79             & 9.28                          & 11.39                 & 9.10              & 4          & 97.2        & 94.1 & 96.5 & 88.5             & 96.6 & 95.3             & 96.6           & 91.2             & 95.6           & 87.8        \\
\textbf{Gen 2} & 672        & 8.65                            & 8.81             & 8.38                          & 11.02                 & 8.23              & 5          & 96.8        & 93.8 & 96.6 & 88.2             & 96.3 & 95.1             & 96.0           & 91.3             & 93.8           & 77.7        \\
\textbf{Gen 3} & 768        & 8.01                            & 8.63             & 8.05                          & 10.55                 & 8.15              & 6          & 97.0        & 93.7 & 96.4 & 87.5             & 95.0   & 94.3             & 94.9           & 88.7             & 92.9           & 73.1        \\
\textbf{Gen 4} & 864        & 7.61                            & 8.54             & 7.81                           & 9.67                 & 8.00              & -            &          &                          &                          &              &                          &              &            &              &            &             \\
\textbf{Gen 5} & 960        & 7.46                            & 8.14             & 7.52                          & 8.98                 & 7.89              & -            &           &                          &                          &              &                          &              &            &              &            &             \\ \bottomrule
\end{tabular}
}
\label{table:results}
\vspace{-3.5mm}
\end{table*}

We evaluated our neural diffing approach on two speech tasks each encapsulating a data evolution scenario described in Section 1. 

\subsection{Speech Recognition}
For our first set of experiments, our objective is a classic speech recognition task.
The goal will be to deploy a series of ASR models, each trained over more data than the previous to simulate a classic ML flywheel setting where as more users interact with a product, the data set continues to expand.
We capture Setting 1 with this experiment, where the underlying ASR task remains constant but training data becomes more abundant at each model release.
We use the LibriSpeech data set, which is a speech corpus of read English audio books \cite{Panayotov2015}.
The data set provides $\sim 960$ hours of training data and we use its ``clean'' test set to report our word error rate (WER).
Our generation 0 model was trained on $480$ hours of data and then we increment the size of the training data with $96$ additional hours (10\%) at each generation.

Our model is a basic unidirectional LSTM RNN-T architecture with 6 transcription network layers and 2 prediction network layers. 
Each LSTM has 1024 units, the prediction network uses a dropout rate of $0.3$, and both networks apply a final projection layer of size 640.
We apply a single joint layer with tanh activation and use a 2500 word-piece vocabulary with a 256 dimensional input embedding.
The resulting model has 64.1M parameters.
Audio feature extraction uses 64 dimension log-filterbank energy with feature frames downsampled by 3 and stacked with a stride size of two to produce 30 millisecond frames.
We also apply SpecAugment masking \cite{Park2020} on training data.
Decoding is conducted using standard RNN-T beam search with a width of 16.

For each generation and compression method, we built baseline models (BLC) to demonstrate the added benefit of using diffing as opposed to attempting to compress a fresh model trained from scratch at the same compression rate.
We applied sparsity and hashing methods at a compression factor of $10\times$ (i.e. objects will be $90\%$ sparse or hash to a $90\%$ smaller array). 
For sparsifying our baseline models, we prune over steps 20k-35k while for diff objects we prune earlier from 1k-11k.
For our hashing, each weight matrix of the models are hashed to weight arrays $\nicefrac{1}{10}$ th their original size with three hash functions, and as mentioned in Section \ref{sec:hash_diff}, we seed the hash functions differently at each generation.
All diff weights are initialized to zero at the start of training across experiments.
The results are shown in Table \ref{table:results} where we also include unconstrained baselines trained from scratch (BL).

One observes that both the sparse and hash diffs outperform their comparable compressed baselines BLC generation over generation.
However, the sparse diffing method consistently maintains nearly matching WERs with BL at each generation (e.g. 7.46 BL vs 7.52 Diff at 960 hours) whereas, hash diffing performance begins to diverge, resulting in a final WER degradation of 8\% relative against the unconstrained ideal BL.
Also note that for our setup Hash BLC had difficulty training to satisfactory levels.
\vspace{-0.7mm}

\subsection{SLU Classification}
\vspace{-0.3mm}
For our second set of experiments, we demonstrate the effectiveness of our diffing approach on a speech command SLU task.
We use the  publicly available Fluent Speech Commands (FSC) \cite{Lugosch2019} dataset to conduct these experiments.
FCS contains $\sim 30$k utterances from $97$ unique speakers and is decomposed into train, dev, and test sets with $\sim 23$k, $3$k, and $4$k utterances respectively.
Each FCS utterance is one of $248$ phrasings, each of which is associated with a domain of intent: \textit{Lighting}, \textit{Heat}, \textit{Volume}, \textit{Bring}, \textit{Language Change}, or \textit{Music}.

The modeling objective for the experiment is one of classification where we aim to correctly identify each utterance as one of a predfined set of supported phrases.
To emulate Setting 2, where the underlying task changes at each generation, we evolve the system by behaving as if the local assistant edge device is being enabled with new functionality releases which need speech support.
We do this with three successive updates.
Namely, our generation 0 model begins by only being trained on and classifying phrases among the \textit{Lighting}, \textit{Heat} and \textit{Volume} domains.
At generation 2, we introduce the \textit{Bring} domain data and add support for its phrases. 
Likewise at generation 3, we add in \textit{Language Change} commands, and finally generation 4 introduces the \textit{Music} domain phrases.
At each generation our accuracy metric is evaluated on the subset of test data encompassing only those supported domains up to that point.

Our model architecture here will be based on a contrastive approach with a speech encoder and text encoder, much like the approach of \cite{Agrawal2020}.
The speech encoder and text encoder jointly learn to embed representations from their modality of input into a shared space where corresponding utterance acoustic and text phrase embeddings are close with respect to the $L_2$ distance while nonmatching are well separated.
We apply the \emph{triplet loss} with online batch mining \cite{Hermans2017} where batches contain audio utterances and the full set of supported text phrases.
At runtime, nearest neighbor search is conducted from the speech encoder embedding against the reference embeddings of the supported phrases generated by the text encoder.
Our speech encoder is a $4$-layer LSTM with $128$ units with a dense projection layer with $128$ units.
We also use the same audio feature extraction as performed with LibriSpeech data.
Our text encoder consists of a $2$-layer LSTM with $128$ units again with a projection output layer, and it also has $64$-dimension input embeddings over a vocabulary of 4500 word pieces.
The resulting architecture has 1.1M parameters.
The recurrent layers are trained with a dropout rate of $0.2$, and we used the Adam optimizer coupled with a simple hold and decay learning rate\footnote{An exception was the hash diffs which were altered to hold a higher rate for longer in some instances and converged before decay.}.
Our sparsity schedule for all sparse training operates over steps 2k-12k.

For this set of experiments, we ran with two different levels of compression/data budgets, $20\times$ and $40\times$.
Because of the relative simplicity of the task, we can apply these higher compression levels, but also, they serve to magnify the effect of the diffing approaches over their baselines.
We also included a static baseline ($\text{BL}_{\infty}$) where we do not update the model at all to show the performance drop at each step.
The results are shown in Table \ref{table:results} using phrase classification accuracy as the measure for model performance.
As with the LibriSpeech experiments, we found the sparse methods for diffing maintain high accuracy at each generation with the most extreme $40\times$ compression only degrading accuracy by 2\% absolute with respect to BL, a 7.5\% improvement over its baseline BLC.
While we see moderate success with hash diffing over baselines BLC, it was not as pronounced as sparsity.
It is also important to note that for this task, hashing achieved better accuracy results for simply compressing the baseline model over sparsity (e.g. $95.1$ vs. $93.8$ for 5 domains) .
Furthermore, during the training process we found hash diffing much more challenging to tune, a lack of consistency across updates, and a more expensive training process (both higher compute time per step and more steps to converge).
The significance of these findings highlight that while a compression method might be a promising technique to directly shrink the footprint of a model trained from scratch, the method might not directly translate into success at compressing a diff.
\vspace{-1.5mm}

\section{Conclusion}
\vspace{-0.5mm}
We address the problem of limited network OTA budgets for transferring generations of speech models from server to device.
We detail two architecture agnostic diff methods based on sparsity and hashing, which learn compact representations to increment a model locally.
We show these methods can compress to diff objects to high degrees, up to $40 \times$ in some cases, without significant impact on predictive performance.
We find that not all compression techniques adapt equally well for diffing, and in future work we would like to see more approaches proposed with additional update mechanisms introduced beyond basic additive and multiplicative schemes.

\bibliographystyle{IEEEtran}

\bibliography{mybib}


\end{document}